# The Art of Planetary Science:
# Art as a Tool for Scientific Inquiry and Public Discourse around Space Exploration

**Jamie L. Molaro**

**Abstract.** Art can be a powerful tool in science engagement efforts to help facilitate learning and public discourse around space and space exploration. The Art of Planetary Science is an annual exhibition combining science and art which aims to help people to connect more meaningfully to science outside of traditional education models. Works solicited from scientists and from the public explore the beauty of the universe, as well as communicate and abstract scientific concepts from an artistic framework. These events offer the public a unique perspective on science and an opportunity to participate in dialogue around how and why we explore space. As an extension of the exhibition, a series of workshops for artists and educators focuses on techniques in creating science-driven art and how it can be used as a tool for scientific inquiry. We will discuss our success with these efforts and the important role that art can play in shaping the evolving narrative of humanity's relationship to space.

**Introduction**
Art is a critical tool for expressing, exploring, and reflecting on complex ideas and draws many parallels to the pursuit of science. It can be leveraged in public outreach efforts[1] to help people engage with science more meaningfully and to facilitate the broad discussion of our relationship to space happening in society today. In the age of climate change, the burgeoning private space industry, and a science literacy crisis, it has never been more important to facilitate critical discourse on role of science in society, how and why we explore space, and who is shaping the narrative in our relationship with space. Towards this end, The Art of Planetary Science (TAPS)[2] is an outreach effort combining science and art at annual exhibitions, along with a companion workshop series Making Space that aims to teach artists and educators how to use art as a tool for scientific inquiry.

---

[1] Aafke Fraaije, Marjoleine G. van der Meij, Frank Kupper, and Jacqueline EW Broerse, 'Art for public engagement on emerging and controversial technologies: A literature review', *Public Understanding of Science,* Vol. 31, no. 6 (2022): pp. 694-710; Catherine Lau, Chantal Barriault, and Julia Krolik, 'Evaluating the impact of a comprehensive Canadian science-art residency program on the participating scientist, artist and the public', *Frontiers in Education*, Vol. 6 (2022): 690489.

    The motivation behind TAPS was to present a different perspective on science to people outside academia. Visual representation of scientific information is often limited to figures in research publications and textbooks. For those of us who have made it our profession, we see beauty in those figures because they have meaning and implications for the way we understand the world around us. They reflect the laws which govern the physical universe, the processes building the diverse landscapes and cloudscapes of other worlds, and the craft involved in the process of discovery. However, without this context, such beauty can be hard to derive. TAPS provides a venue that allows people to explore science through a flexible artistic framework, and over the years we have observed how greater access to this framework can enrich and transform individuals in a fundamental way. As a society, we tend to think of the role of scientists as creating knowledge and of artists as creating self-expression, but both are pursuits of exploration and inquiry and, together, they become a powerful tool for learning, discourse and reflection. We will discuss our success with these efforts and the important role that art can play in science engagement around space and space exploration.

**The Art of Planetary Science**
TAPS is an annual art exhibition at the University of Arizona's Lunar and Planetary Laboratory, founded by graduate students (led by the author) in 2013. The event features fine art with subjects inspired by astronomy and the Solar System, alongside works that incorporate or communicate scientific ideas or data. In the latter case, scientists are encouraged to create art using the data or concepts involved in their own research and other participants are encouraged to learn and explore new concepts through their art. Displaying both types of works together creates a unique perspective on space, and a vertex for interdisciplinary dialogue between the science and art communities and the general public.

---

[2] Jamie L. Molaro and James T. Keane, 'The Art of Planetary Science: An Exhibition – Bringing Together the Art and Science Communities to Engage the Public', *45th Annual Lunar and Planetary Science Conference* (2014): No. 1397.

While realism and scientific illustration have historically played an important role in the perception of science[3], participants creating science-driven works are also encouraged to abstract, reinterpret, and reshape information, allowing the audience to engage with it in a less rigid framework and leveraging the opportunity to show science in a new light. The most compelling works tend to be ones which do not try to communicate too much, but instead lead the viewer to a simplified concept and allow them to experience its complexity to whatever degree they choose. For example, *Exo-Resonance* by Dr. Josh Lothringer (Fig. 1) visualizes the intangible by depicting orbital resonances between planets and their stars. Captured in a repeating dance by gravity, the line between two planets in one of these special orbits draws out an intricate geometrical pattern reflecting cosmic order. In *Spiral Moon*, Dr. JA Grier (Fig. 2) plays with shade and contrast using a three-dimensional technique called paper quilling to explore the concepts of albedo and the opposition effect, in which the brightness of the Moon's surface changes with the materials it is made of and with viewing angle.

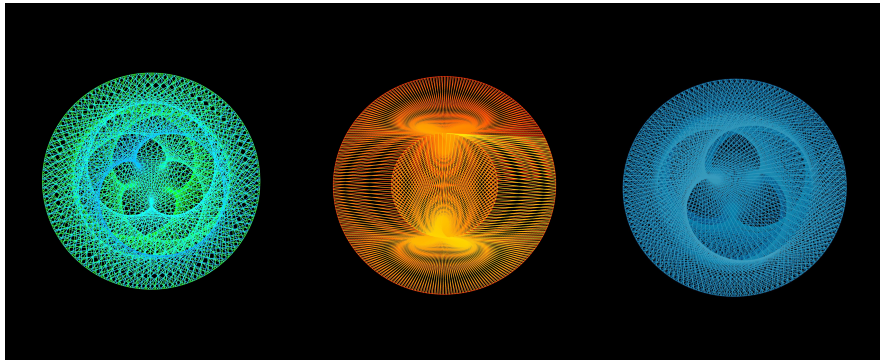

Figure 1. *Exo-Resonance*, digital illustration, Josh Lothringer

---

[3] Lois R. Rosson, 'The Astronomical Realists: The Social Mechanics of Visual Documentation, Art, and the American Space Age, 1944–1987', (Ph.D. dissertation, University of California Berkeley, 2022); L. Rosson and R. Miller, 'Space Art as a Modern Movement: From the Moon to Today', in John Ramer and Ron Miller, eds., *The Beauty of Space Art*, 2nd ed. (Springer International Publishing, 2021), pp. 73-91.

8  The Art of Planetary Science: Art as a Tool for Scientific Inquiry and
   Public Discourse around Space Exploration

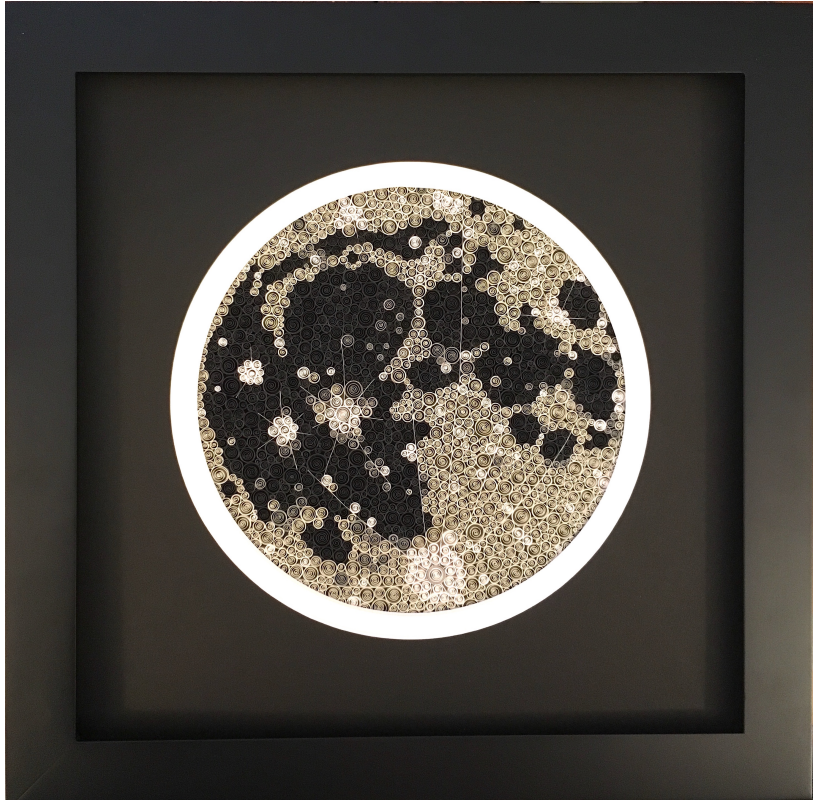

Figure 2. *Spiral Moon*, paper, JA Grier

The first exhibition in 2013 featured seventy works from nearly fifty artists, drawing a few hundred attendees to the one-night opening. Encouraged by its success, the event was extended to a full weekend the following year and by its third had grown so much that we were forced to start jurying artwork due to space constraints. A typical event now features 220-250 works of art from 100-120 individual artists, some new each year and some returning, and sees an audience of at least several hundred. A wide range of mediums are represented, including painting, drawing, photography, textiles, paper, ceramics, sculpture, digital works, and poetry. Video submissions facilitate performance-based mediums like music and dance, as well as animations. Participants are primarily from the local

Tucson community and the greater southwestern United States, but some works are shipped from throughout the country (even internationally). Over the years, collaborations with other university departments and local museums have yielded special features such as the cross-display of art, participatory art interventions, and live music performances. Since its founding, TAPS has organized fifteen exhibitions including both annual and satellite (below) events.

Maximizing engagement with science requires removing barriers to participation, and to this end inclusion is a core philosophy of TAPS. Anyone regardless of background is encouraged submit art and participants come from a wide range of professions and skill levels, from hobbyists to gallery-selling artists. Due to its popularity, some jurying must occur each year, however allowing artists to submit multiple pieces enables us to cut submissions rather than participants. This philosophy is key because it helps remove inhibitions that level of skill or knowledge should limit access to or exploration of science itself. If the aim of the event is explicitly to encourage interdisciplinary inquiry and dialogue, then no individual can be required to have expertise in both (or either) professions[4]. We also try to broaden our reach by holding smaller satellite events in conjunction with planetary and astronomy conferences. These events solicit works from artists local to where the conferences are held, allowing us to reach new communities and providing artists an opportunity to share their work with a large international audience. We recommend this approach to others organizing an art-related event for the first time because it carries lower cost and risk than a standalone show, since the venue and audience are already provided by the conference.

**The Role of Art in Space Exploration**
From the standpoint of science outreach, these events have been highly successful at engaging new audiences with space exploration, complementing existing efforts in space science and astronomy, such as citizen science, science festivals, and informal K-12 education[5]. TAPS fills

---

[4] Sarah Davies, Ellen McCallie, Elin Simonsson, Jane L. Lehr and Sally Duensing, 'Discussing dialogue: Perspectives on the value of science dialogue events that do not inform policy', *Public understanding of science 18,* No. 3 (2009): pp. 338-353; Jay Lemke, Robert Lecusay, Michael Cole, and Vera Michalchik. *Documenting and assessing learning in informal and media-rich environments*. (The MIT Press, 2015).

[5] Bruno Strasser, Jérôme Baudry, Dana Mahr, Gabriela Sanchez, and Elise Tancoigne. '"Citizen science"? Rethinking science and public participation',

a niche by reaching a primarily adult audience and members of the art community. While we do not collect quantitative outcomes, our feedback surveys indicate that participants and audience alike learn and experience a sense of enrichment from engaging with science through art, and from the opportunity to meet the artists and scientists presenting works. Of course, a relevant question is *how substantive or superficial is this enrichment*? Through observation and interaction with participants and audiences over the years, we have come to understand that art has a tremendous power to facilitate interdisciplinary learning and societal discourse in ways that can profoundly impact both the public and the science community itself, which we attempt to articulate here.

*(1) Creating and/or engaging with science- and data-driven art provides the opportunity to form a meaningful, personal connection to science outside the confines of a classroom.*

Public engagement is critical to the field of planetary science to share the wonder of other worlds, promote interest in STEM education, and garner public support for funding NASA programs. This is accomplished through efforts such as press events and classroom materials, through which we have become adept at generating public excitement about space exploration. Unfortunately, it is much more challenging to get people excited about the science that drives that exploration and we often fail to communicate the details of our research effectively. Part of this challenge stems from the fact that science is something many people believe is beyond their ability to comprehend. This belief is often ingrained at a young age by the reinforcement of implicit biases and societal misconceptions about who can succeed in STEM, lack of exposure to scientists as "real people" and role models, and demographic factors such as gender, race, and socioeconomic status[6]. This problem is exacerbated by

---

*Science & Technology Studies,* Vol. 32, no. 2 (2019): pp. 52-76; Magdalena Kersting, Rolf Steier, and Grady Venville. 'Exploring participant engagement during an astrophysics virtual reality experience at a science festival', *International Journal of Science Education, Part B* 11, no. 1 (2021): pp. 17-34; Nuala O'Flynn. 'Cosmic Curiosity: Why collaborating with libraries results in impactful family engagement with astronomy', *Communicating Astronomy with the Public Journal*, Vol. 32 (2023): pp. 9-14.

[6] Jiyun Elizabeth L. Shin, Sheri R. Levy, and Bonita London, 'Effects of role model exposure on STEM and non-STEM student engagement', *Journal of*

Jamie L. Molaro   11

the fact that most scientific research is only communicated to other scientists, and many interesting discoveries are never reported on by the media and remain obscured in technical journals. As long as people believe that science is beyond their ability to understand, they will feel alienated by it, and that alienation can have profound impacts on our society as a whole[7].

The Art of Planetary Science was created to help directly address this issue. Art connects us emotionally and experientially to its subject beyond language[8]. By exhibiting art containing scientific ideas or created from

---

*Applied Social Psychology,* Vol. 46, no. 7 (2016): pp. 410-427; Jessica R. Gladstone and Andrei Cimpian, 'Which role models are effective for which students? A systematic review and four recommendations for maximizing the effectiveness of role models in STEM', *International Journal of STEM Education,* Vol. 8, no. 1 (2021): pp. 1-20; Eleanor K. Chestnut, Ryan F. Lei, Sarah-Jane Leslie, and Andrei Cimpian, 'The myth that only brilliant people are good at math and its implications for diversity', *Education Sciences,* Vol. 8, no. 2 (2018): 65; Sapna Cheryan, Allison Master, and Andrew N. Meltzoff, 'Cultural stereotypes as gatekeepers: Increasing girls' interest in computer science and engineering by diversifying stereotypes', *Frontiers in Psychology,* Vol. 6 (2015): 49; Lin Bian, Sarah-Jane Leslie, Mary C. Murphy, and Andrei Cimpian, 'Messages about brilliance undermine women's interest in educational and professional opportunities', *Journal of Experimental Social Psychology,* Vol. 76 (2018): pp. 404-420; Elizabeth Litzler, Cate C. Samuelson, and Julie A. Lorah. 'Breaking it down: Engineering student STEM confidence at the intersection of race/ethnicity and gender', *Research in Higher Education,* Vol. 55 (2014): pp. 810-832; Stephen Gorard and Beng Huat See, 'The impact of socio-economic status on participation and attainment in science', *Studies in Science Education,* Vol. 45, no. 1 (2009): pp. 93-129.

[7] Kosuke Motoki, Toshiki Saito, and Yuji Takano, 'Scientific literacy linked to attitudes toward COVID-19 vaccinations: a pre-registered study', *Frontiers in Communication,* Vol. 6 (2021): 707391; Bastien Trémolière and Hakim Djeriouat 'Exploring the roles of analytic cognitive style, climate science literacy, illusion of knowledge, and political orientation in climate change skepticism', *Journal of Environmental Psychology,* Vol. 74 (2021): 101561.

[8] Dahlia W. Zaidel, 'Art and brain: insights from neuropsychology, biology and evolution', *Journal of Anatomy 216,* No. 2 (2010): pp. 177-83; Beth Kessler, *Picturing the cosmos: Hubble Space Telescope images and the astronomical sublime* (University of Minnesota Press, 2012); Christopher W. Tyler, and Lora T. Likova, 'The role of the visual arts in enhancing the learning process', *Frontiers in Human Neuroscience,* Vol. 6 (2012): 8.

scientific data, it provides the opportunity for people to form a meaningful, personal connection to science in a way that does not require them to "learn" it[9]. There are no classrooms or tests, and the experience between individual and art is private which allows them to engage without fear of judgement. Importantly, this experience is also voluntary[10], allowing them to choose the depth to which they explore the content. One individual may only skim an artwork description while another may be eager to learn more. Both experiences achieve our aim; the model values connection over learning because establishing the former can lead to increased interest in the latter during future opportunities to engage with science[11], and because "learning" takes many forms beyond facts, including identity development[12]. Hosting the events in the university department also welcomes artists into science spaces and explicitly makes their contributions part of the environment. This helps to erode the barrier which makes people feel excluded from academia and which separates science and art as distinct pursuits. Even for individuals who never move beyond connection, an increased feeling of comfort and confidence engaging with science and scientists can transform the way they think about it in a fundamental way, and perhaps have implications for things like how they vote, lifestyle choices, and their ability to critically analyze information. At its broadest level, this has implications for the way that society grows as a result of the creation of scientific knowledge.

---

[9] Davies, 'Discussing dialogue', (2009): pp. 338-353.

[10] Lemke, *Documenting and assessing learning*. (2015); John H. Falk and Lynn D. Dierking, 'Reimagining public science education: the role of lifelong free-choice learning', *Disciplinary and Interdisciplinary Science Education Research,* Vol. 1 (2019): pp. 1-8.

[11] Eric Jensen and Nicola Buckley. 'Why people attend science festivals: Interests, motivations and self-reported benefits of public engagement with research' *Public Understanding of Science 23,* No. 5 (2014): pp. 557-573; John H. Falk, Scott Pattison, David Meier, David Bibas, and Kathleen Livingston, 'The contribution of science-rich resources to public science interest', *Journal of Research in Science Teaching,* Vol. 55, no. 3 (2018): pp. 422-445.

[12] Lemke. *Documenting and assessing learning*. (2015); Davies, 'Discussing dialogue', (2009): pp. 338-353.

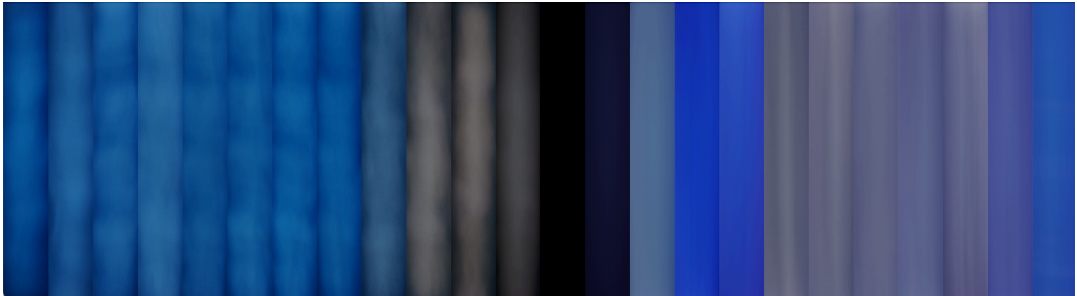

Figure 3. *24 Hours*, print on canvas, Cui Jing

While we do not collect formal survey data from these events, this power of art to connect is evident through informal interactions with artists who tell us TAPS is the first time they have made art about space, and especially with those creating science-driven works despite having no formal science background. One artist who participates each event shared with us that she is retired and only paints a single painting each year, specifically for our show. It is also evident in interactions with audience members who seek out scientists during the event with questions and ideas, by the audience the event darws, some of whom express having little to no previous interaction with the science community, and by the recurring participation of many artists and audience members over multiple years[13].

*(2) As a tool, art offers new ways to explore ideas and worlds, leading to new understanding and perspectives.*

The opportunity to learn from science-driven works occurs on a spectrum for both the audience and the creators, but especially for the latter. An artist's job is to use their craft to explore and express ideas, and even for scientific topics it can be a powerful practice of inquiry. For example, in *24 Hours* (Fig. 3), Cui Jing photographed a piece of white paper on the ground each of twelve hours a day in Tucson, Arizona, USA where she lived and in Hangzhou, China where she is from. Their times zones are 15 hours apart and thus, together, the series of 24 images represents a full day in her life. This work not only reflects her personal relationship to her environment but is itself a scientific observation of the way that light, atmospheric conditions, and air quality interact to change the appearance and color of the sky differently across Earth's surface. Ekaterina Smirnova explored Comet 67P/Churyumov–Gerasimenko in a series of paintings (Fig. 4) by using electrolysis to create her own deuterium ("heavy water", which comets contain) with which to paint its unusual landforms, and she invented a water vaporization technique to recreate the spray of particles caused by ice sublimation at its surface. To make the art more immersive, a digital app that she used to scan the paintings provides soundscapes and information about the comet's composition, and postcards allow the audience to experience its unpleasant but complex smell. Such works may not explore or communicate concepts by the same method as a professional scientist would use, but instead offer a completely different way of relating to and understanding them.

---

[13] Davies, 'Discussing dialogue', (2009): pp.338-353.

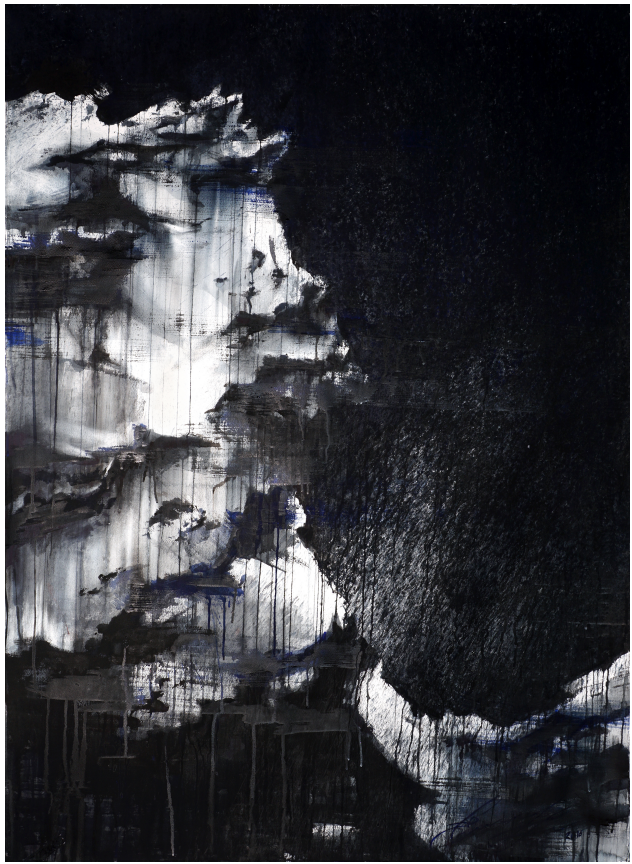

Art can bring new perspective to professional scientists as well. Science is an inherently creative practice in that requires imagination and interpretation. Learning begins with collecting observations, but the data itself cannot tell us what it means. It requires imagination to make the cognitive leap to turn information into understanding, and the way you interpret that information determines where you leap. However, an individual's creativity in these acts can be both expanded and limited by their provenance.

Figure 4. *67P XII*, water-media on paper, Ekaterina Smirnova

Due to the nature of academia, the demographics and types of people who have the most access to it, and the traditional ways in which scientists are trained, many scientists have similar approaches to research and problem solving and therefore may miss what more a diverse, transdisciplinary perspective could bring to their work[14].

In a personal anecdote, lunar scientist, painter, and co-founder of the International Association of Astronomical Artists (IAAA) Dr. William K. Hartmann recalled an encounter at a scientific conference about comets in the 1980s. Hartmann had done a painting imagining the surface geology of a comet nucleus and asked various experts at the meeting for critiques about its depiction. However, comet surfaces are usually hidden by the coma of dust and gas being emitted from the nucleus as it nears the Sun, and therefore are typically not observable through telescopes. Hartmann was surprised to find that most comet experts who viewed the painting had never given much thought to what the surface might look like, having instead only focused their study on what they could see, and it spurred much discussion amongst them.

Such an example illustrates how artists are good at asking questions from new and different perspectives. In this way, empowering artists to use art to explore scientific knowledge can directly enable them to participate in building it. Science as a practice can benefit by encouraging such participation, and scientists themselves can become better researchers by learning from the more flexible frameworks from which they approach a subject[15]. The act of creating art around scientific topics thus facilitates scientific progress from both sides.

*(3) As a form of self-expression, art allows people outside academia to participate in defining how and why society values and benefits from scientific knowledge.*

---

[14] Jay H. Bernstein, 'Transdisciplinarity: A review of its origins, development, and current issues', *Journal of Research Practice,* Vol. 11 (2015); Katherine W. Phillips, Douglas Medin, Carol D. Lee, Megan Bang, Steven Bishop, and D. N. Lee, 'How diversity works', *Scientific American* Vol. 311, no. 4 (2014): pp. 42-47.

[15] Chris Impey and Dinah Jasensky, 'Visualizing the Universe – The Intersection of Art and Astronomy', *Mediterranean Archaeology and Archaeometry 18,* no. 4 (2018): pp. 307-318.

16  The Art of Planetary Science: Art as a Tool for Scientific Inquiry and
    Public Discourse around Space Exploration

Figure 5. *Genesis*, colored pencil on paper, Mica Post

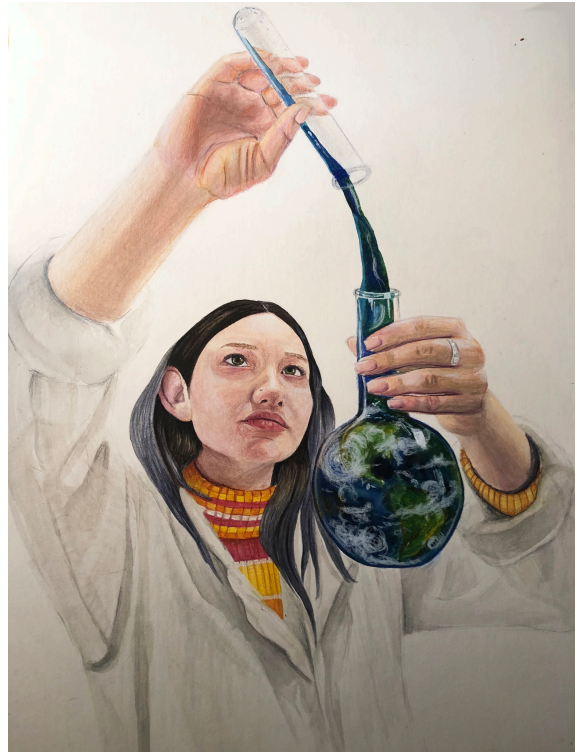

As scientists, we are the keepers of our body of knowledge. As such, it is natural to center engagement efforts around what we personally find inspiring about space as if, by virtue of it being our profession, we can define this for all people. However, it is critical that we reverse the conversation and look at our work from other points of view. On one hand, a person cannot be inspired by something they do not know and therefore we must inevitably make choices about how and what we present. Further, since the public does not have our deep understanding of science, part of our role is to help guide society in using that knowledge to benefit humanity. On the other hand, the content we choose and viewpoint we take in presenting our work to the public will naturally reflect both explicit and implicit biases and beliefs held by the scientific community, which is not representative of society as a whole. It is crucial that all voices be included in the dialogue[16] around how we use and obtain scientific knowledge, and the ways in which society can be enriched and grow from it.

    Art offers a medium through which some laypeople may more easily express a point of view about science and society, and through which they

---

[16] Davies, Sarah R. 'Constructing communication: Talking to scientists about talking to the public.' *Science communication 29*, Nno. 4 (2008): pp. 413-434; Davies, 'Discussing dialogue', (2009): pp. 338-353.

can be encouraged to participate in this dialogue. By asking them to create art focused around space, TAPS creates an environment which invites their input on important questions facing us in the modern era: Should we explore space? Where should we explore? Should we send robotic or crewed missions? Should we work to protect or to utilize it as a natural resource? What do we envision for our future relationship to space? How do we face the challenges of living on our own planet?

Diverse viewpoints can be seen in the wide range of topics artists choose to explore at TAPS events. Some are straightforward representations of where or what participants are passionate about or wish for humanity's future, while others are more philosophical, critical, or questioning. For example, *Genesis* by Mica Post (Fig. 5) depicts a woman in a lab coat in an apparent act of creation, holding a beaker which contains the Earth. The power of such a piece comes from the many ways it might be interpreted: Does it depict God as a scientist or humanity as creators? It could call attention to how we alter our environment or reflect a desire to understand the origin of life. Perhaps it is a self-portrait, and it is her passion for science which is seen emerging. My own work, Margaret's Moon (Fig. 6), explores a narrative of knowledge building in a three-dimensional cratered surface of layered paper, cut using topography data collected by the Lunar Reconnaissance Orbiter spacecraft. The printed text is computer code from the Apollo lunar command module, juxtaposing the knowledge we created to land on the Moon with data collected from the science program it enabled. The title refers to Margaret Hamilton who led the development of the code, commenting on who historically has been allowed to participate in space exploration and what contributions we may be missing due to lack of diversity in the field.

**Making Space: A Workshop on Space, SciArt, & Society**[17]

As an extension of the exhibitions, a workshop series aims to facilitate interdisciplinary learning and discourse more directly and intentionally with artists and educators, helping them engage more deeply with science content and giving them the tools to use art in learning and teaching. The workshops also emphasize the principals behind TAPS and role they personally can play in space exploration. By doing so, they serve not only to cultivate a connection to science and dialogue among the participants themselves, they also empower participants to do so for others. By incorporating the ideas and techniques learned into their own works and careers, they will expose their own audiences to art as a way of learning and facilitate dialogue on space and society in new contexts and communities.

---

[17] Jamie L. Molaro, Sanlyn R. Buxner, Amanda Hendrix, and Alex Parker, 'Making Space: A Workshop on Space, Sci-Art, and Society', *NASA Exploration Science Forum* (2022).

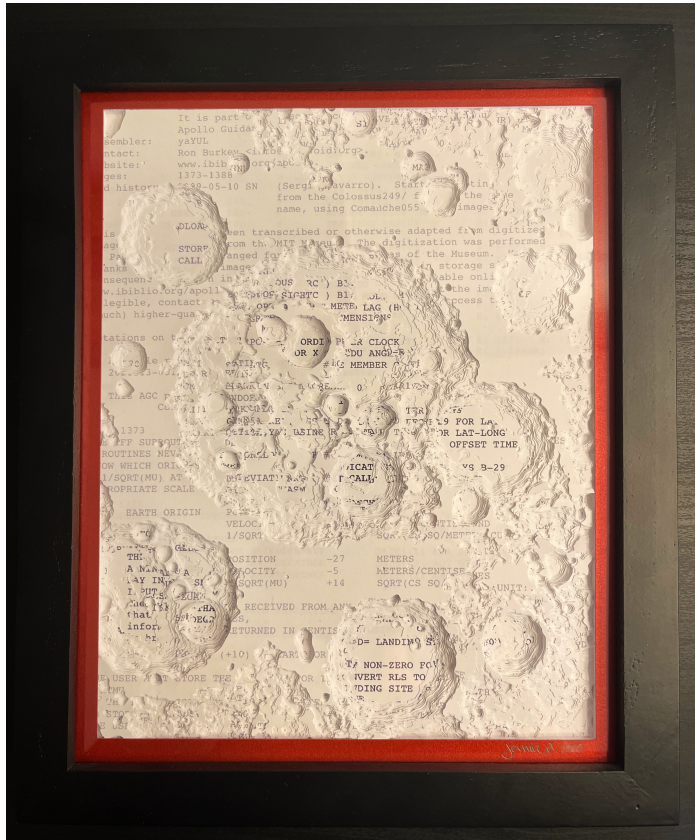

Figure 6. *Margaret's Moon*, paper, Jamie Molaro

Making Space is a multi-day workshop centered around using art to explore scientific concepts and humanity's relationship to space. Theme talks focus on connections between science and art as practices, the perception of science in society, techniques in creating science-driven art, the history of space art, and accessing public data from space missions. Science talks cover topics in planetary science, with emphasis on how spacecraft observations help us learn about other worlds. Small group discussions encourage dialogue about our relationship to space and the purpose and value of exploration, paired with collaborative art-making prompts which draw connections between exploration and creativity. The event is targeted towards an adult audience with some background (professional or otherwise) in art and/or science, including artists, scientists, engineers, science writers, educators, and more.

   Approximately half of the workshop is spent doing parallel solo activities called "labs," analogous to science courses in higher education. Labs are directed art-making activities that model approaches to incorporating scientific ideas or data into a work of art and in using the artistic process to learn. For example, in one lab participants use an online planetarium to collect classification information about their favorite constellation, learning about different attributes of stars which they represent visually in the artwork. In another, they use paint to explore how images of gas giants taken in different wavelengths differ and what it reveals about their clouds. Other activities ask participants to abstract an idea into music or poetry, or tell an exploration narrative in origami or clay. The goal of the labs is to enjoy and practice different ways of blending science and art that may spur inspiration for individuals in their own mediums. At the end of the workshop, participants are asked to create their own piece of science-driven art using what they have learned and present it to the group remotely.

   The pilot workshop was held in April 2022 in Tucson, Arizona, USA with 25 participants, and a second in July 2023 in Pasadena, California, USA with 29 participants. In the former, approximately half the participants had backgrounds in art, a quarter in science or engineering, and a third in education or the humanities, while the breakdown of the latter was approximately half in art, half in science or engineering, and only a few in education or humanities. Survey questions provided to both groups resulted in a response rate of ~44%, or a total of 24 responses. Participants were asked: 1) *Has this workshop increased your interest in learning about space or science?* 2) *Has this workshop made learning about space or science seem more accessible to you?* 3) *In what ways has this experience changed how you think about art and/or science, the ways you might pursue either, or your identity as an artist and/or scientist?* During the second workshop, a question was added: 4) *Has this workshop increased your interest in participating in space exploration discourse through art?*

       In response to question 1, 83% of surveys indicated increased interest in learning about space and science. This included participants with science backgrounds, some of whom noted interest in learning about new

20  The Art of Planetary Science: Art as a Tool for Scientific Inquiry and Public Discourse around Space Exploration

areas of science beyond their research topics. The remaining 17% of respondents either did not answer the question or indicated their interest could not be increased any more than it currently was. In response to question 2, 79% of surveys indicated feeling that space and science seemed more accessible for one of two primary reasons. Some respondents cited better knowledge of available resources and techniques for finding or using information and data from space missions than prior to the workshop. Others indicated increased confidence, either in their ability to learn about scientific topics or in how well accepted their interest in doing so via art might be accepted by their professional peers. For question 3, major themes conveyed by respondents were a shift in their perspective on science, better understanding of how art and science relate in their own practices, and intent to pursue more science-driven art in the future. Of these, 35% of respondents indicated some kind of change in self-perception or broadening of their self-identity as an artist or scientist to include aspects of the other, and an additional 13% indicated their self-identity, which already included both, was affirmed. Of respondents from the second workshop, 88% indicated increased interest in participating in discourse about space exploration, while the remaining 12% responded that they did not feel prepared to do so. Approximately half of participants created their own artworks to present to the group in the weeks following the workshop, indicating some longevity in their engagement with the content, and a few have kept longer term contact providing updates on space-related artwork they have created. The following quotes are examples of feedback we received either verbally or from the surveys:

'As an artist I've always felt alienated by science even though I was interested in it. But now I feel like you've broken down barriers I had put up for myself and I can interact with the world in a completely new way.' –*Violet Brand, artist (Tucson 2022)*

'I discovered a yearning to learn more about space and science. I even have the thought to lean into science when I work towards my master's degree. It can be intimidating at times, but this workshop showed just how accessible it can be.' –*Anonymous, artist (Pasadena 2023)*

'I think I might try using art as a way to meditate on and progress my own research.' –*Anonymous, scientist (Tucson 2022)*

'This lab really shows how art can be used as a different way of learning something. I want to try using it in my class.' –*Dr. Michelle Wooten, astronomy educator*

"I came away challenged to do more and say more with my art."
–*Danielle Rose, artist (Pasadena 2023)*

**Parting Thoughts**
The ability for events like these to recontextualize science outside of an academic setting is an extremely powerful tool. The first quote above, in particular, demonstrates how transformative it can be for a person to access and engage with science from within their own framework and when that access has previously been limited. Even within traditional classroom settings, art can be an accessible way to engage students in learning[18], as demonstrated by Dr. Wooten (fourth quote) who reported success in implementing a workshop lab in her undergraduate astronomy class. We are encouraged by the success of our efforts over the years and heartened by the positive impact we have observed it can have on participants and attendees.

   The idea of putting science and art together is not new, and our work builds on top of those artists and communicators coming decades, even centuries, before us.[19] Art was what brought space into modern public imagination, led by painters like Chesley Bonestell and Lucien Rudaux in the early 20th century. Others like Carl Sagan and the artists of the IAAA helped bring the discoveries of the growing United States space program to public eye throughout its latter half.[20] Today, new images from space telescopes are eagerly anticipated by the public[21] and science fiction in art and media abounds. Yet, as space becomes an ever-bigger part of life on Earth, society's perception of and relationship to it will continue to evolve and it is critical that discourse surrounding it does too. We seek to grow beyond past efforts to connect individuals more directly with space and science through their active participation in making and engaging with art, helping to enrich their relationship to the universe and give them a voice in how we explore it.

---

[18] Tyler, 'The role of the visual arts in enhancing the learning process', (2012); Ayça Akın, 'Let me make mathematics and music together: A meta-analysis of the causal role of music interventions on mathematics achievement', *Educational Studies* (2023): pp. 1-19.

[19] Impey and Jasensky, 'Visualizing the Universe'*,* pp. 307-318.

[20] Rosson, 'The Astronomical Realists', (2022); Rosson and Miller, 'Space Art as a Modern Movement', pp. 73-91.

[21] Kessler, *Picturing the cosmos* (2012).

**Acknowledgements**

Development of the Making Space workshop was supported by the NASA Solar System Exploration Research Virtual Institute 2016 (SSERVI16) through nodes Project ESPRESSO and TREX under the cooperative agreements 80ARC0M0008 and NNH16ZDA001N, respectively. Thanks go to the artists who agreed to include their works in this paper, the authors of the quotes and anecdotes, and to my co-founders from the early days of TAPS, Dr.'s James Keane, Sarah Peacock, and Hannah Tanquary.